\documentclass[acmtog]{acmart}
\usepackage{amsfonts}       
\usepackage{amsmath}       
\usepackage{siunitx}
\usepackage{colortbl}       
\usepackage{xcolor}         

\usepackage{acmart-taps}
\newcommand{\myrowcolour}{\rowcolor[gray]{0.95}}
\newcommand{\grey}[1]{\textcolor{gray}{#1}}

\acmSubmissionID{0569}

\AtBeginDocument{%
  }

\acmDOI{10.1145/3588432.3591533}
\acmISBN{979-8-4007-0159-7/23/08}

\begin{document}

\title{Neural Parametric Mixtures for Path Guiding}

\author{Honghao Dong}
\affiliation{%
  \institution{Peking University}
  \city{Beijing}
  \country{China}}
\email{cuteday@pku.edu.cn}
\orcid{0000-0001-7247-1301}

\author{Guoping Wang}
\affiliation{%
  \institution{Peking University}
  \city{Beijing}
  \country{China}}
\email{wgp@pku.edu.cn}
\orcid{0000-0001-7819-0076}

\author{Sheng Li}
\authornote{Corresponding author.\\ 
Project website: \href{https://neuropara.github.io}{https://neuropara.github.io}.}
\affiliation{%
  \institution{Peking University}
  \city{Beijing}
  \country{China}}
\email{lisheng@pku.edu.cn}
\orcid{0000-0002-8901-2184}

\begin{abstract}
   Previous path guiding techniques typically rely on spatial subdivision structures to approximate directional target distributions, which may cause failure to capture spatio-directional correlations and introduce parallax issue. In this paper, 
    we present Neural Parametric Mixtures (NPM), a neural formulation to encode target distributions for path guiding algorithms. 
    We propose to use a continuous and compact neural implicit representation for encoding parametric models while decoding them via lightweight neural networks. 
    We then derive a gradient-based optimization strategy to directly train the parameters of NPM with noisy Monte Carlo radiance estimates. 
    Our approach efficiently models the target distribution (incident radiance or the product integrand) for path guiding, and outperforms previous guiding methods by capturing the spatio-directional correlations more accurately.
    Moreover, our approach is more training efficient and is practical for parallelization on modern GPUs. 
\end{abstract}



\maketitle

\section{Introduction}
\label{sec:introduction}
The efficiency of path tracing relies heavily on the sampling strategy. To further improve its efficiency and robustness, path guiding algorithms leverage the knowledge gained during rendering to facilitate the process of light-path construction, thereby reducing noise. 
To acquire better importance sampling distribution, local path guiding techniques employ previous radiance estimates to learn an approximation of spatial incident radiance fields, which are then used to guide the construction of paths. In practice, current methods typically use some representation (e.g., Gaussian mixtures \cite{vorba2014line, herholz2016product}, quadtrees \cite{muller2017practical}) to approximate the directional distribution of incident radiance. A spatial subdivision structure (e.g., kd-tree \cite{ dodik2022path}, or octree \cite{Bus2017DoubleHierarchies}) is then used to store these distributions, thus accounting for the spatial variations. 

However, several key deficiencies remain in their paradigm. 
Most methods learn the marginalized incident radiance distribution within each subdivided spatial region. This fails to capture the spatio-directional correlations within the spatial discretizations, and could cause artifacts (e.g., parallax error, Fig~\ref{fig:parallax}(a)). 
Moreover, their spatial subdivision structures are subject to frequent reconstruction for finer-grained spatial resolution, which needs extra overhead and require a long training time to converge.
Meanwhile, it is challenging to efficiently fit these specific directional distributions from noisy samples, especially in an online manner \cite{ruppert2020robust}.

While an adaptive and robust spatial representation is difficult to achieve with manually designed subdivision schemes, we saw the recent success of neural implicit representation in compactly modeling spatially varying functions with fine-grained and high-frequency details~\cite{mildenhall2020nerf}.
In this work, we exploit the great expressiveness of neural implicit representation while preserving the desirable properties of parametric mixture models (e.g. efficient importance sampling) for path guiding algorithms. We thereby present Neural Parametric Mixtures (NPM), which use a continuous and compact implicit representation to encode spatio-directional target distributions, and decode them into PMMs with lightweight neural networks for fast importance sampling. We show that our NPM representation, without explicit spatial subdivision schemes, can be efficiently trained simply using gradient-based optimization techniques. Specifically, our method has advantages in the following aspects: 

First, our continuous implicit representation of spatial radiance fields naturally captures the correlations between spatial positions and directional target distributions. 
By smoothly interpolating and decoding the implicit representations with neural networks, our method inherently avoids the issues due to spatial discretization, thus resulting in higher performance.  

Second, our compact representation avoids the extra overhead and long training time caused by the iterative reconstruction strategies applied to the explicit spatial subdivision structures. Combined with our simple optimization based on stochastic gradient descent, our method outperforms other guiding methods even with fewer training samples. In addition, our method is practical and performant for parallelization on GPU.

Lastly, our method can learn the product distribution (i.e., multiplied by the BSDF and the cosine term). This further reduces the noise with a modest computational overhead while not requiring the extra effort of previous solutions (e.g., fitting each BSDF with pre-computed parametric models).

\section{Related Work}
\label{sec:related}

\paragraph{Path Guiding} 
To achieve better sampling strategies, local path guiding techniques leverage previous radiance estimates (either online or during a pre-computation process) to build an approximation of the incident radiance fields, which is used to guide subsequent sampling. Early approaches used simple bases such as histograms for importance sampling, e.g. built from a photon map \cite{jensen1995importance} or collected radiance estimates with 5-D tree structures \cite{lafortune19955d}. Subsequent work has developed various techniques to construct the guiding distribution, e.g., Gaussian mixtures \cite{vorba2014line}, quad-trees \cite{muller2017practical}, which is often stored in spatial data structures (e.g., kd-tree and octree) to account for spatial variations of the distributions.

Deep learning techniques have also been explored recently, achieving improvements while often with less practical performance. 
For example, convolutional networks could be used to reconstruct the learned noisy radiance field \cite{huo2020adaptive,zhu2021hierarchical}.
Specially designed neural networks could also model complex manifolds \cite{dinh2016density}, while allowing samples to be drawn directly from the learned distribution \cite{muller2019neural}. However, the prohibitive computational cost prevents its practical application \cite{muller2019neural, vorba2019production}. Instead of directly importance sampling using neural networks, we encode the target distribution into implicit neural representation, and use only lightweight MLPs to decode it into parametric mixtures for efficient sampling. We show that our method can be efficiently trained ($< 10$s per scene on a single GPU) while being sufficiently robust and practical.

\paragraph{Parametric Mixture Models}
Parametric mixture models (PMMs) are convex combinations of parametric distributions, and are often used to approximate directional distributions in graphics applications. They have many desirable properties, e.g., fast sampling, and closed-form solutions for products, convolutions and integrals. Several types of PMMs (e.g., Gaussian mixtures \cite{vorba2014line, dodik2022path} and von Mises-Fisher mixtures \cite{ruppert2020robust}) are widely used in the recently developed path guiding algorithms.
Several recent works also use PMMs to fit BSDFs with precomputation \cite{ruppert2020robust, herholz2016product}, and multiply them with the learned incident radiance to achieve product sampling.

Parametric models can also be predicted by neural networks, enabling new possibilities for e.g. lighting \cite{currius2020spherical} and reconstruction \cite{yu2021plenoctrees} tasks. In this work, we use neural representations to encode parametric mixtures for efficient sampling. Our method is also naturally extensible to product sampling.

\paragraph{Implicit Neural Representation}
Following the success of using neural networks to represent 3D scenes implicitly \cite{mildenhall2020nerf}, the concept of neural representation has been popularized and applied to various tasks. They use sparse input images to optimize the spatial radiance fields via a differentiable volume rendering procedure, thus enabling novel view synthesis. Inspired by its recent successful applications \cite{diolatzis2022active, muller2022instant}, we exploit a continuous and compact implicit neural representation to encode the spatio-directional target distributions for path guiding algorithms. While the ground truth target distribution (i.e., the incident radiance or product distribution) is unknown, our NPM representation can be optimized online using minibatch stochastic gradient descent (SGD), where the gradients for training are estimated by Monte Carlo integration using noisy radiance estimates. 

\section{Preliminary}
\label{sec:background}

\paragraph{Monte Carlo Integration}
Light transport algorithms are generally based on the rendering equation \cite{kaj86re}:
\begin{equation}
L_{\mathrm{o}}\left(\mathbf{x}, \omega_{\mathrm{o}}\right)=L_{\mathrm{e}}\left(\mathbf{x}, \omega_{\mathrm{o}}\right)+\int_{\Omega} f_{\mathrm{s}}\left(\mathbf{x}, \omega_{\mathrm{o}}, \omega_{\mathrm{i}}\right) L_{\mathrm{i}}\left(\mathbf{x}, \omega_{\mathrm{i}}\right)\left|\cos \theta_{\mathrm{i}}\right| \,\mathrm{d} \omega_{\mathrm{i}},
\label{eq:re}
\end{equation}
which defines the relationship between the outgoing radiance $L_{\mathrm{o}}$, emitted radiance $L_e$, and the integrated incident radiance $L_{\mathrm{i}}$, at shading point $\mathbf{x}$. Monte Carlo integration is used to obtain an estimate of the reflection integral $L_r$ using an average of $N$ samples. In the case where $N=1$:
\begin{equation}
\left\langle L_{\mathrm{r}}\left(\mathbf{x}, \omega_{\mathrm{o}}\right)\right\rangle= \frac{f_{\mathrm{s}}\left(\mathbf{x}, \omega_{\mathrm{o}}, \omega_{\mathrm{i}}\right) L_{\mathrm{i}}\left(\mathbf{x}, \omega_{\mathrm{i}}\right)\left|\cos \theta_{\mathrm{i}}\right|}{p(\omega_{\mathrm{i}} \mid \mathbf{x}, \omega_{\mathrm{o}})},
\end{equation}
where $\langle L_{\mathrm{r}}\left(\mathbf{x}, \omega_{\mathrm{o}}\right)\rangle$ is an unbiased estimate of the outgoing radiance $L_{\mathrm{r}}\left(\mathbf{x}, \omega_{\mathrm{o}}\right)$, and $\omega_i$ is the incident direction sampled with some directional probability distribution $p(\omega_{\mathrm{i}} \mid \mathbf{x}, \omega_{\mathrm{o}})$. The variance of this estimator $V[\langle L_{\mathrm{r}}\rangle]$ can be reduced if the sampling distribution resembles the shape of the integrand, and could even reach zero variance if being proportional to it (i.e., $p \propto f_s\cdot L_{\mathrm{i}}\cos \theta_i$). This, however, is difficult to achieve with only BSDF importance sampling, leaving the remaining part of the integrand (i.e., the incident radiance) unknown, resulting in a relatively high variance of the MC estimator. Path guiding algorithms, on the other hand, manage to obtain better importance sampling strategies often by using previous radiance samples to approximate the incident radiance $L_{\mathrm{i}}$ or the full integrand $f_s\cdot L_{\mathrm{i}}\cos \theta_i$, which will be discussed later.

\paragraph{Von Mises-Fisher Mixtures} We use the \emph{von Mises-Fisher} (vMF) distribution as the basis of NPM. The vMF distribution is defined as:
\begin{equation}
v(\omega \mid \mu, \kappa)=\frac{\kappa}{4 \pi \sinh \kappa} \exp \left(\kappa \mu^{T} \omega\right),
\end{equation}
where $\mu \in \mathbb{S}^2$ and $\kappa \in [0, +\infty)$ defines the direction and precision (sharpness) of the vMF distribution. The vMF mixture model (VMM) is thus a convex combination of $K$ vMF components/lobes:
\begin{equation}
\mathcal{V}(\omega \mid \Theta)=\sum_{i=1}^{K} \lambda_{i} \cdot v\left(\omega \mid \mu_{i}, \kappa_{i}\right),
\end{equation}
where $\Theta$ contains the parameters ($\mu_i$, $\kappa_i$) and weights ($\lambda_i)$ of each vMF component. The vMF mixtures have many desirable properties, e.g., fewer parameters (4 floats per component), efficient importance sampling, and closed-form product and integration, which together constitute the reason for choosing it as the basis of NPM. 

Our key is to encode the vMF mixtures with our implicit neural representation, then decode them with lightweight MLPs, and train them to effectively model the target distributions for path guiding algorithms. 
Other parametric basis functions (e.g., Gaussian mixtures) could be integrated into our method using a similar paradigm. 

\section{Neural Parametric Mixtures}
\label{sec:method}

\begin{figure*}[!ht]
  \centering
  \includegraphics[width=\linewidth]{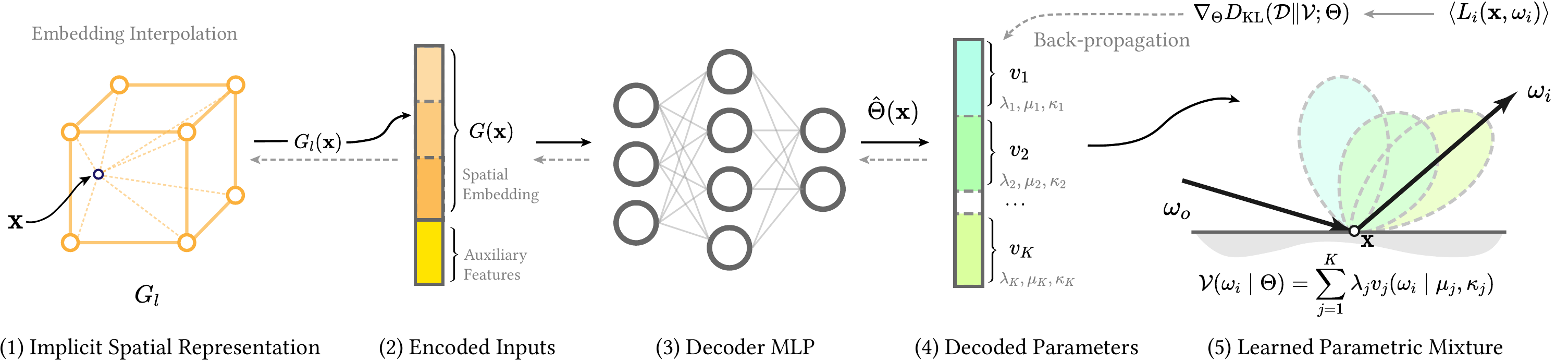}
  \caption{
  High-level illustration of our Neural Parametric Mixtures (NPM). We implicitly encode the spatially varying target distributions with the multi-resolution embedding. When the distribution of a spatial location $\mathbf{x}$ is queried, \textbf{(1)} the features assigned to the nearby grid points surrounding $\mathbf{x}$ are interpolated at each level, and concatenated with other levels to obtain the spatial embedding $G(\mathbf{x})$. \textbf{(2)} the spatial embedding is then combined with other inputs to \textbf{(3)} feed into the lightweight MLP for \textbf{(4)} decoding the parameters $\Theta$ of the vMF mixture $\mathcal V (\omega_i \mid \Theta)$ with $K$ components. We then \textbf{(5)} use this parametric distribution for importance sampling the scattering direction. The result MC radiance estimate $\langle L_{\mathrm{i}}(\mathbf{x}, \omega_i) \rangle$ is used to estimate the training gradient $\nabla_\Theta D_{\mathrm{KL}}$ (Sec.~\ref{ssec:trainnpm}), which is then back-propagated through these differentiable stages to optimize our NPM representation (dashed lines).
  \label{fig:pipeline}}
\end{figure*}

In this section, we present our Neural Parametric Mixtures (NPM) technique for local path guiding. We first show how to encode/decode target distributions with NPM in a simple setup (i.e., learning incident radiance fields, Sec.~\ref{ssec:radnpm}), then we derive the optimization method for NPM based on minibatch stochastic gradient descent (Sec.~\ref{ssec:trainnpm}). Finally, we show how our NPM could naturally benefit from learning the full integrand (to account for the BSDF term), as well as the other extensions for better learning target distributions (Sec.~\ref{ssec:extendnpm}).
An overview of our method is illustrated in Fig.~\ref{fig:pipeline}.

\subsection{Radiance-based NPM}
\label{ssec:radnpm}
\begin{figure}[!t]
  \centering
  \includegraphics[width=\linewidth]{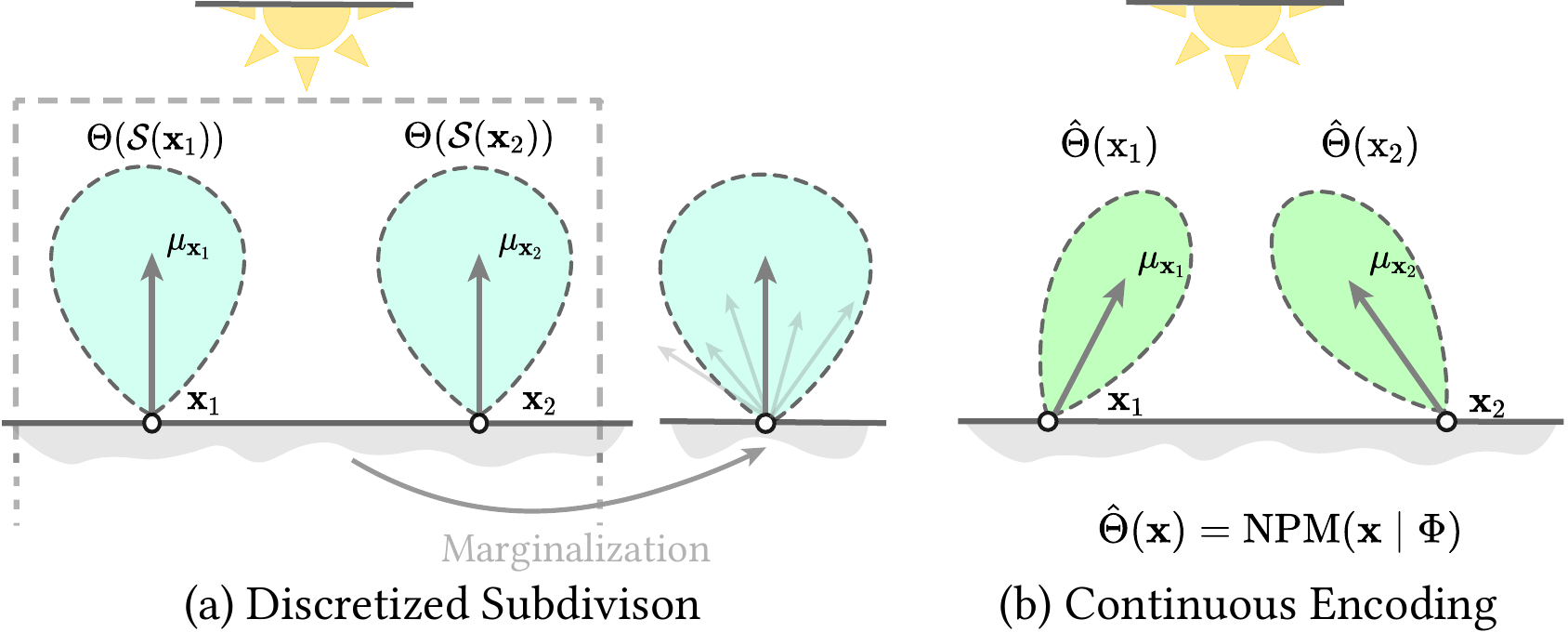}
  \caption{
  Parallax issue caused by spatial discretizations (a). For a subdivided volume $\mathcal S(\mathbf{x})$ in (a), the guiding distribution is marginalized with training samples scattered over the volume $\mathcal S(\mathbf{x})$, and is shared by different positions (e.g., $\mathbf{x}_1$ and $\mathbf{x}_2$). Our method will not suffer from parallax due to NPM implicitly representing a monolithic function, continuously mapping from spatial positions to parametric guiding distributions, as shown in (b).
  \label{fig:parallax}}
\end{figure}

In order to acquire a better importance sampling strategy, we should obtain an approximation of the incident radiance distribution using previous radiance estimates, known as the radiance-based local path guiding \cite{herholz2016product, rath2020variance}. Specifically, we want to use the vMF mixtures to be approximately proportional to the incident radiance, at a given shading position $\mathbf{x}$:
\begin{equation}
    \mathcal{V}\left(\omega_i \mid \Theta\left(\mathbf{x}\right)\right) \propto L_{\mathrm{i}} (\mathbf{x}, \omega_i),
\label{eq:guidedist}
\end{equation}
where $\Theta$ is conditioned on $\mathbf{x}$ to account for the spatial variation of the target distribution.
Previous work achieves this with specific spatial subdivision strategies (e.g., kd-tree, octree). 
However, this spatial discretization introduces artifacts (e.g., resulting from parallax, Fig.~\ref{fig:parallax} (a)), and is subject to frequent reconstruction to converge to a fine grained spatial subdivision, as discussed in Sec. \ref{sec:introduction}. 

Instead, we use an implicit neural representation to encode the target distribution compactly. This allows the spatial variation of the distribution to be continuously accounted for, thus better capturing spatio-directional correlations. Technically, given a shading position $\mathbf{x}$ in the scene, our NPM would output the guiding distribution that approximates the target distribution (Eq.~\ref{eq:guidedist}). The output guiding distribution is defined using a set of parameters $\hat{\Theta}(\mathbf{x})$: 
\begin{equation}
    \mathbf{NPM}(\mathbf{x} \mid \mathbf{\Phi}) = \hat{\Theta}(\mathbf{x}),
    \label{eq:npmradiance}
\end{equation}
where $\Phi$ are the trainable parameters of the implicit representation, and $\hat{\Theta}$ are the output decoded parameters, defining a vMF mixture $\mathcal{V}(\omega_i \mid \hat\Theta(\mathbf{x}))$ that is trained to approximate $L_{\mathrm{i}}(\mathbf{x}, \omega_i)$ (Eq.~\ref{eq:guidedist}). By continuously conditioning the learned distribution $\Theta$ on spatial positions $\mathbf{x}$,  our method inherently avoids the above issues caused by spatial discretizations. We achieve the above mapping by using a lightweight network to decode this parametric distribution from the implicit neural representation. To make sure that we get a valid vMF mixture (i.e., $\lambda_i, \kappa_i > 0$, $\mu_i \in \mathbb {S}^2$, and $\sum_{j=1}^K \lambda_j = 1$), we must additionally regularize the raw network output with appropriate mapping functions (see Tab.~\ref{tab:activations}). Specifically, we apply exponential activation to $\lambda_i$ and $\kappa_i$. Logistic activation is applied to $\theta_i$ and $\varphi_i$, which form the spherical coordinates of $\mu_i$. Most importantly, we apply the softmax function to all $\lambda$s to ensure that the outputs model a valid PDF (i.e., satisfy $\sum_{i=1}^{K} \lambda_i = 1$).

\begin{table}[!b]
    \centering
    \caption{Detailed mapping functions we use to regularize network outputs, where $\lambda^{\prime}$, $\kappa^{\prime}$ $\theta^{\prime}$, $\varphi^{\prime}$ denote the raw outputs, and $(\theta, \varphi)$ is the normalized spherical coordinate of $\mu \in \mathbb {S}^2$. Left: parameter notations and their valid ranges; middle: type of activation; right: specific mappings. }
    \label{tab:activations}
    \begin{tabular}{ccc} 
        \toprule
        {Parameter} & {Activation} & {Mapping} \\ 
        \midrule
        $\kappa \in [0, +\infty)$ & \text{Exponential} &  $\kappa_i = \exp( \kappa_i^{\prime})$ \\
        $\lambda \in [0, +\infty)$ & Softmax &  $\lambda_i = \exp (\lambda_i^{\prime}) / \sum_{j=1}^{K} \exp (\lambda_j^{\prime})$ \\
        $\theta$\text{, } $\varphi \in [0, 1]$  & Logistic  & $\theta_i = 1 / (1 + \exp (- \theta_i^{\prime}))$ \\
        \bottomrule
    \end{tabular}
\end{table}

\paragraph{Discussion.} It is possible to implement different forms of implicit neural representation with trainable parameters $\Phi$. While it is straightforward to use a monolithic network to model $\mathbf{NPM}_{\Phi}: \mathbf{x} \to \Theta$, we find it difficult to fit the high-frequency variations of the target distribution. Thereby, we use a trainable multi-resolution spatial embedding for encoding the distributions, and additionally a lightweight neural network for decoding the parameters. This is crucial for our method to achieve better modeling capacity while remaining performant, as will be discussed later. 

\subsection{Optimizing NPM}
\label{ssec:trainnpm}
We show how to optimize the divergence between the decoded distribution $\hat \Theta(\mathbf{x})$ and the target distribution using minibatch stochastic gradient descent. To achieve this, the gradients of a training objective (or loss function) with respect to the network parameters are necessary. However, it is non-trivial to define such a loss function, given the 
ground truth output parameters $\Theta _{\mathrm{gt}}(\mathbf{x})$ are unknown. Previous works typically use design optimization algorithms (e.g., expectation-maximization) that iteratively use batches of samples to fit a given set of parameters $\Theta$, which often parameterize a marginalized distribution shared by the spatial region covering the samples~\cite{herholz2016product, ruppert2020robust}. 
However, their methods are applied to explicitly parameterized models, and are therefore not applicable to our method, which models the implicit representation of the function $ \mathbf{NPM}_{\Phi}: \mathbf{x} \to \hat\Theta$.  

We minimize the $\mathrm{KL}$ divergence between the decoded vMF mixtures and the target distribution via minibatch stochastic gradient descent, where its gradients with respect to the trainable parameters are estimated using Monte Carlo integration. Other divergence metrics are also available following a similar derivation. 
Let us start by assuming that the shading position $\mathbf{x}$ is fixed, thus omitting the dependency of $\Theta$ on $\mathbf{x}$ in the equations. For a given position, the $\mathrm{KL}$ divergence between the target distribution $\mathcal D$ and our output distribution $\mathcal{V}$ is defined as:
\begin{equation}
D_{\mathrm{KL}}(\mathcal D \| \mathcal{V}; \Theta) =\int_{\Omega} \mathcal D(\omega) \log \frac{\mathcal D(\omega)}{\mathcal{V}(\omega \mid \hat\Theta)} \,\mathrm{d} \omega,
\end{equation}
where $\mathcal D \propto L_{\mathrm{i}}$ in radiance-based path guiding. 
This integral could now be estimated with the Monte Carlo estimator with $N$ samples:
\begin{equation}
D_{\mathrm{KL}}(\mathcal{D} \| \mathcal{V}; \Theta) \approx \frac {1} {N} \sum^N_{j=1} 
\frac{ \mathcal{D}(\omega_j)}{ \Tilde{p}(\omega_j \mid \hat\Theta)}  \log \frac{\mathcal{D}(\omega_j )}{\mathcal{V}(\omega_j \mid \hat\Theta)},
\end{equation}
where $\tilde{p}$ is the distribution from which the samples are drawn, which in our case is a combination of the BSDF importance sampling and guiding distribution.
By taking its derivative with respect to $\Theta$, we obtain the MC estimate of the gradient $\nabla_{\Theta} D_{\mathrm{KL}}(\mathcal{D} \| \mathcal{V}; \Theta)$:
\begin{equation}
\nabla_{\Theta} D_{\mathrm{KL}}(\mathcal{D} \| \mathcal{V}; \Theta) \approx - \frac {1} {N} \sum^N_{j=1} 
\frac{ \mathcal{D}(\omega_j) \nabla_{\Theta} \mathcal{V}(\omega_j \mid \hat\Theta) }{ \Tilde{p}(\omega_j \mid \hat\Theta) \mathcal{V}(\omega_j \mid \hat\Theta) }  ,
\label{eq:mcgradients}
\end{equation}
where the derivatives of the vMF mixtures $\mathcal V$ with respect to their parameters $\Theta$ are straightforward.
The gradients for the trainable NPM parameters $\Phi$ could then be obtained via back propagation. Since we use the unbiased MC estimate of the training gradients, the parameters are guaranteed to converge to a local minimum. 

In practice, our training sample pairs $(\mathbf{x}, \omega_i) \to L_{\mathrm{i}}$ are distributed in different spatial positions $\mathbf{x}$, efficiently learning a spatially varying target distribution $\mathcal D ( \mathbf{x})$. This results in the training objective accounting for the divergence of multiple positions. The expected solution for $\Phi$ is thus:
\begin{equation}
\Phi^{*} = \underset{\Phi}{\arg \min} \, \mathbb E_{\mathbf{x}} \left[D_{\mathrm{KL}}\left(\mathcal{D}\left(\mathbf{x}\right) \| \mathcal{V}; \Theta \left(\mathbf{x}\right)\right)\right] .
\end{equation}
For our implicit spatial embedding (i.e., grids of latent features, discussed later), this results in the embedding being optimized with all (and only) its nearby samples. When using the gradient descent method, the samples with the largest gradients (i.e., the most important ones for reducing divergence) would dominate, forming a reasonable design choice for better adaptivity. 

\subsection{Full Integrand Learning}
\label{ssec:extendnpm}
Using path guiding to sample the full integrand $f_s\cdot L_{\mathrm{i}}\cos \theta_i$ can achieve even better performance, which should incorporate the BSDF term and the cosine term into the target distribution. This is challenging since the guiding distribution is now conditioned on $\mathrm{5D}$ inputs (i.e., outgoing direction $\omega_{\mathrm{o}}$ and spatial coordinate $\mathbf{x}$). Previous works fit BSDFs with precomputed parametric models and multiply them with the learned incident radiance distribution to achieve product sampling. However, this often relies on scene-dependent precomputation, discretization over $\omega_{\mathrm{o}}$, and extra computational overhead \cite{herholz2016product,ruppert2020robust}.

Our neural design can naturally handle the conditions with the extra input of $\omega_i$. 
This is essential since a neural network could approximate arbitrary conditional models if being expressive enough.  
We later show this improves performance through learning a better guiding distribution, with only modest performance overhead. 
For clarity, we denote the previous radiance-based method as NPM-radiance, and this version as NPM-product.

Specifically, by supplementing input $\omega_{\mathrm{o}}$, we reformulate the learned distribution (Eq.~\ref{eq:npmradiance}) with the outgoing directions. This enables learning the full integrated as:
\begin{equation}
\mathbf{NPM}_{\mathrm{product}}(\mathbf{x}, \omega_{\mathrm{o}} \mid \mathbf{\Phi}) = \hat{\Theta}(\mathbf{x}, \omega_{\mathrm{o}}),
\label{eq:npmproduct}
\end{equation}
where $\hat\Theta$ now parameterizes the vMF mixture $\mathcal V$ that is trained to approximate the full integrand in Eq.~\ref{eq:re}, i.e.,
\begin{equation}
\mathcal{V}\left(\omega_i \mid \hat\Theta \left(\mathbf{x}, \omega_{\mathrm{o}}\right)\right) \propto 
f_{\mathrm{s}}\left(\mathbf{x}, \omega_{\mathrm{o}}, \omega_{\mathrm{i}}\right) L_{\mathrm{i}}\left(\mathbf{x}, \omega_{\mathrm{i}}\right)\left|\cos \theta_{\mathrm{i}}\right|,
\label{eq:guidedistfull}
\end{equation}
where the cosine term could be approximated with a constant vMF lobe \cite{ruppert2020robust}, leaving NPM to focus on the remaining part of the integral. Nonetheless, it is still challenging for neural networks to model a 2D directional distribution conditioned on 5D spatio-directional inputs. We further use the following simple extensions to help the network learn these spatially varying distributions:

\paragraph{Auxiliary Feature Inputs.}
Following the practices in prior work \cite{muller2021real, hadadan2021neural}, we additionally input the surface normal and roughness as auxiliary features to help the network better correlate the target distribution with e.g., local shading frame (normal) and spatially varying BSDFs (roughness). Experimentally, we find this helps the network to better capture the spatio-directional correlations, while with a small computational overhead due to additional memory traffic.

\paragraph{Input Encoding.}
It is challenging for a neural network to model the non-linearity between multidimensional inputs and outputs, especially when our outputs are distributions with high-frequency spatial variations.
Therefore, we replace the spatial input $\mathbf{x}$ with our trainable multi-resolution spatial embedding (discussed in Sec.~\ref{ssec:encoding}). For the other inputs (e.g., outgoing direction $\omega_{\mathrm{o}}$ and surface normals $\mathbf{n(x)}$), we encode them using the spherical harmonics basis, which is previously established in NeRF \cite{verbin2022refnerf}.

\section{Implementation}
\label{sec:impl}
In this section, we provide the technical details that are crucial to the performance and practicality of our NPM implementation.

\subsection{Multi-resolution Spatial Embedding}
\label{ssec:encoding}
Our implicit NPM representation learns a continuous mapping $ \mathbf{NPM}_{\Phi}: \mathbf{x} \to \hat\Theta$ (with the additional input $\omega_{\mathrm{o}} \in \mathbb{S}^2$ in the extended version), where $\Theta \in \mathbb{R}^{4\times K}$ defines the learned target distribution. While a straightforward solution would be using a multi-layer perceptron (MLP) as the universal function approximator to model $ \mathbf{NPM}_{\Phi}$, we experimentally found it difficult to capture the high-frequency spatial variations of the target distributions.

Therefore, we use a learnable spatial embedding to implicitly encode the learned parametric mixtures. Similar approaches are found successful in recent NeRF-like applications \cite{munkberg2021nvdiffrec, muller2022instant}. Specifically, we define $L$ 3D uniform grids $G_l$, each covering the entire scene with a spatial resolution of $D_l ^3$, where $G_l$ denotes the $l\text{-th}$ 
embedding grid. $D_l$ grows exponentially, resulting in multiple resolutions of the embedding. We then assign a learnable embedding (a latent feature vector $v \in \mathbb{R}^F$) to each lattice point of $G_l$. To query the spatial embedding for $\mathbf{x}$, we bilinearly interpolate the features \emph{nearby} $\mathbf{x}$ for each resolution, and concatenate them to obtain the final embedding $G(\mathbf{x})$.
More formally:
\begin{equation}
    G(\mathbf{x} \mid \Phi_{\mathrm{E}}) = \overset{L}{\underset{l = 1}{\oplus}} \mathrm{bilinear} \left(\mathbf{x}, V_l\left[\mathbf{x}\right]\right), \,G: \mathbb{R}^3 \to \mathbb{R}^{L\times F},
\end{equation}
where $V_l[\mathbf{x}]$ is the set of features at the 
eight corners of the cell enclosing $\mathbf{x}$ within $G_l$. 
The spatial embedding $G(\mathbf{x})$ is then concatenated with other inputs (e.g., $\omega_{\mathrm{o}}$ and auxiliary features) to the MLP for decoding the parameters $\Theta$.
We thus formulate the desired mapping (taking Eq.~\ref{eq:npmradiance} for example) as a two-step procedure:
\begin{equation}
    \mathbf{MLP}\Bigl(G(\mathbf{x} \mid \Phi_{\mathrm{E}}) \bigm\vert \Phi_{\mathrm{M}} \Bigr) = \hat \Theta (\mathbf{x}),
\end{equation}
where the parameters of the spatial embedding ($\Phi_{\mathrm{E}}$) and the MLP ($\Phi_{\mathrm{M}}$) together constitute the trainable parameters $\Phi$ of our implicit representation for NPM.
Intuitively, a spatial embedding implicitly encodes the target distribution within a specific spatial region, while the multi-resolution design efficiently accounts for different levels of detail (LOD).
By smoothly interpolating between the spatial embedding around positions and decoding them using neural networks, we naturally account for the spatial variations of the target distribution.  
This also lessens the burden of using a single monolithic MLP as the implicit representation, leaving it mainly focusing on decoding it into parametric models $\Theta$. This significantly accelerates training/inference with a larger memory footprint.

\subsection{Online Training Scheme}
\label{ssec:training}

\paragraph{Renderer Integration}
We implement our method on a custom GPU-accelerated renderer based on OptiX \cite{parker2010optix}, where the training and inference procedures are integrated into a \emph{wavefront}-style path tracer \cite{laine2013megakernels}. This design choice allows ray casting, importance sampling, and BSDF evaluation to be performed in coherent chunks over large sets of traced paths by splitting the traditional megakernel path tracer into multiple specialized kernels. This improves GPU thread utilization by reducing the control flow divergence.
Most importantly, this allows us to efficiently sample and evaluate the guiding distributions at each vertex along the path in parallel, thus significantly accelerating network training/inference.

Specifically, we place the training/inference samples into \emph{queues}, where the structure-of-arrays (SoA) memory layout is applied to improve memory locality. At each ray intersection of the chunk of traced paths, the queries for guiding distributions within the queue are processed via batched network inference. The sampling and evaluation procedures are then performed, also using specialized kernels, before entering the next ray-cast kernel. This provides our method with maximum parallelism through large-batch training and inference, minimizing the latency caused by waiting network queries, while avoiding inefficient single-sample inference. 

\begin{figure*}[!t]
  \centering
  \includegraphics[width=\linewidth]{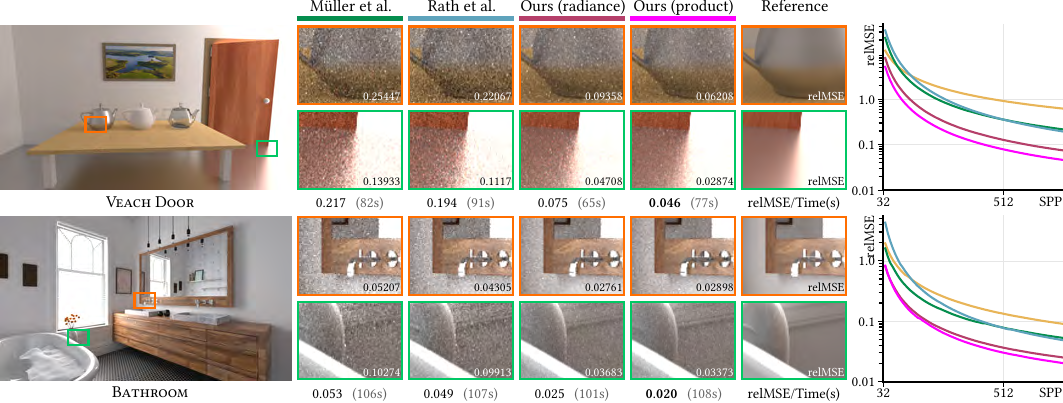}
  \caption{
  Equal-sample-count ($\text{750spp}$) comparisons for two scenes. We show the error (for both the zoom-in areas and whole images) and time cost of different methods. The yellow plots (as well as the other figures) refer to the results obtained by unidirectional path tracing. 
  \label{fig:core_comp}}
\end{figure*}

\paragraph{Training Scheme}
We use the same configuration to train each scene online during rendering, without any scene-specific fine-tuning or pre-computation.
During training, we collect MC radiance estimates along each traced path, and split them into mini-batches for training. The optimization step is performed for each spp, which allows drawing samples to be drawn from the latest guiding distribution. The distribution of the samples (for both rendering and training) is thus gets refined as training proceeds. We stop the training process after a fixed fraction of the total rendering budget (either time or sample count). While we always set this to $25\%$ in our experiments, we find our NPM technique converges quickly during training, generally reaching a local minimum after about $150\text{spp}$, which amounts to about 1000 training steps/batches and 15s (including the runtimes of both training and rendering) on GPU.

\subsection{Guiding Network}
\label{ssec:architecture}
We implement our network on the \emph{tiny-cuda-nn} framework \cite{muller2021tcnn} and integrate it into our renderer. The MLP we used (for both $\text{NPM-{radiance}}$ and $\text{NPM-{product}}$) contains 3 linear layers of width $64$. Each layer with ReLU activation, except for the last layer with our custom mapping functions (Tab.~\ref{tab:activations}). We let the network output $K=8$ vMF components, i.e., $\Theta \in \mathbb {R}^{8 \times 4}$. For the multi-resolution spatial embedding, we use $L=8$ grids with increasing resolutions for each level. The coarsest level has a resolution of $D_1 = 8$ while the finest level has $D_8 = 86$. The feature of each level contains $F=4$ floats, resulting in the final spatial embedding $G(\mathbf{x}) \in \mathbb {R}^{8 \times 4}$.
In practice, we find that the performance of the network could be improved by enlarging the capacity of the MLP or the spatial embedding, leaving this a trade-off between quality and speed.  

For training, we use a fixed learning rate of $0.005$ that is large enough to acquire a fast convergence speed. Adaptive momentum techniques like Adam \cite{kingma2014adam} are used for more robust training and better convergence. For importance sampling the decoded mixtures, we use the numerically stable strategy for vMF \cite{jakob2012numerically}. When inference, we also apply exponential moving average (EMA) to the weights of previous training steps, which better reduces the noise of the MC estimated gradients (Eq.~\ref{eq:mcgradients}).  

\section{Results and Discussion}
\label{sec:results}
We run all the experiments on an Intel Core i9-11900 CPU and an NVIDIA RTX3070 GPU. Following the similar practices of previous works \cite{muller2019path, rath2020variance}, we disable NEE and Russian roulette for all methods and set the maximum path length to 10. All methods are implemented upon a GPU path tracing renderer.

We render all images at the resolution of $1280\times 720$, and evaluate image quality using mean relative squared error (relMSE). All the images, additional metrics (MAPE and MRSE), and the false-color maps can be interactively inspected with our supplementary viewer. 

\aptLtoX[graphics=no,type=html]{
\begin{table*}[!htb]
    \centering
    \caption{Practical Path Guiding (PPG)~\cite{muller2019path}, Variance-aware Path Guiding~\cite{rath2020variance}, unidirectional path tracing and our method on 10 test scenes. 
    We report relMSE, render time, and speedup using PPG as the baseline. 
    Our NPM technique consistently reduces the error in the test scenes.} 
    \label{tab:comparisons}
    \begin{tabular}{lccccccccccccccl} 
        \toprule
        & \multicolumn{2}{c}{} 
        & \multicolumn{3}{c}{\cite{muller2019path}} 
        & \multicolumn{3}{c}{\cite{rath2020variance}} 
        & \multicolumn{6}{c}{Ours} \\ 
        \cmidrule(lr){4-6} \cmidrule(lr){7-9} \cmidrule(lr){10-15}       
        & \multicolumn{2}{c}{PT (BSDF)}	
        & \multicolumn{3}{c}{PPG (baseline)}
        & \multicolumn{3}{c}{Variance. PG}
        & \multicolumn{3}{c}{NPM (radiance)}
        & \multicolumn{3}{c}{NPM (product)} \\
        \midrule
        \myrowcolour
        Bathroom & 0.0905 & \grey{48s} & 0.0530 & 1.0 $\times$ & \grey{106s} & 0.0485 & 1.09 $\times$ & \grey{107s} & 0.0251 & 2.11 $\times$ & \grey{101s} & \textbf{0.0203} & 2.61 $\times$ & \grey{108s} \\

        Bedroom & 0.0383 &  \grey{40s} & 0.0201 & 1.0 $\times$ & \grey{105s} & 0.0161 & 1.26 $\times$ & \grey{109s} & 0.0150 & 1.35 $\times$ & \grey{84s} & \textbf{0.0146} & 1.38 $\times$ & \grey{90s} \\
        \myrowcolour
        Breakfast Room & 0.0094 & \grey{48s} & 0.0069 & 1.0 $\times$ & \grey{100s} & 0.0047 & 1.46 $\times$  & \grey{103s} & 0.0038 & 1.80 $\times$ & \grey{63s} & \textbf{0.0035} & 1.96 $\times$ & \grey{71s} \\
        
        Living Room & 0.0273 & \grey{32s} & 0.0184 & 1.0 $\times$ & \grey{74s} & 0.0146 & 1.26 $\times$ & \grey{80s} & 0.0157 & 1.17 $\times$ & \grey{47s} & \textbf{0.0132} & 1.39 $\times$ & \grey{54s} \\
        \myrowcolour
        Pink Room & 0.0046 & \grey{37s} & 0.0082 & 1.0 $\times$ & \grey{74s} & 0.0061 & 1.34 $\times$ & \grey{76s} & 0.0033 & 2.42 $\times$ & \grey{53s} & \textbf{0.0026} & 3.21 $\times$ & \grey{62s} \\
        
        Salle de Bain & 0.0819 & \grey{38s} & 0.0223 & 1.0 $\times$ & \grey{116s} & 0.0346 & 0.64 $\times$ & \grey{116s} & 0.0196 & 1.14 $\times$ & \grey{79s} & \textbf{0.0140} & 1.59 $\times$ & \grey{86s} \\ 
        \myrowcolour
        Staircase & 0.1812 & \grey{34s} & 0.0298 & 1.0 $\times$ & \grey{80s} & 0.0261 & 1.14 $\times$ & \grey{86s} & 0.0194 & 1.54 $\times$ & \grey{72s} & \textbf{0.0172} & 1.74 $\times$ & \grey{76s} \\
        
        Veach Door & 0.6208 & \grey{33s} & 0.2167 & 1.0 $\times$ & \grey{82s} & 0.1945 & 1.11 $\times$ & \grey{91s} & 0.0750 & 2.89 $\times$ & \grey{65s} & \textbf{0.0461} & 4.69 $\times$ & \grey{77s} \\      
        \myrowcolour
        Veach Egg & 8.2918 & \grey{33s} & 0.8379 & 1.0 $\times$ & \grey{82s} & 0.7870 & 1.07 $\times$ & \grey{85s} & 0.5984 & 1.40 $\times$ & \grey{62s} & \textbf{0.5352} & 1.56 $\times$ & \grey{69s} \\ 
        White Room & 0.0301 & \grey{38s} & 0.0278 & 1.0 $\times$ & \grey{107s} & 0.0253 & 1.10 $\times$ & \grey{103s} & 0.0124 & 2.25 $\times$ & \grey{76s} & \textbf{0.0100} & 2.75 $\times$ & \grey{87s} \\ 
        \bottomrule
    \end{tabular}
\end{table*}
}{
\begin{table*}[!htb]
    \centering
    \caption{Practical Path Guiding (PPG)~\cite{muller2019path}, Variance-aware Path Guiding~\cite{rath2020variance}, unidirectional path tracing and our method on 10 test scenes. 
    We report relMSE, render time, and speedup using PPG as the baseline. 
    Our NPM technique consistently reduces the error in the test scenes.} 
    \label{tab:comparisons}
    \begin{tabular}{lccccccccccccccl} 
        \toprule
        & \multicolumn{2}{c}{} 
        & \multicolumn{3}{c}{\cite{muller2019path}} 
        & \multicolumn{3}{c}{\cite{rath2020variance}} 
        & \multicolumn{6}{c}{Ours} \\ 
        \cmidrule(lr){4-6} \cmidrule(lr){7-9} \cmidrule(lr){10-15}
        
        & \multicolumn{2}{c}{PT (BSDF)}
        & \multicolumn{3}{c}{PPG (baseline)}
        & \multicolumn{3}{c}{Variance. PG}
        & \multicolumn{3}{c}{NPM (radiance)}
        & \multicolumn{3}{c}{NPM (product)} \\
        \midrule
        \myrowcolour
        \textsc{Bathroom} & 0.0905 & \grey{48s} & 0.0530 & 1.0 $\times$ & \grey{106s} & 0.0485 & 1.09 $\times$ & \grey{107s} & 0.0251 & 2.11 $\times$ & \grey{101s} & \textbf{0.0203} & 2.61 $\times$ & \grey{108s} \\

        \textsc{Bedroom} & 0.0383 &  \grey{40s} & 0.0201 & 1.0 $\times$ & \grey{105s} & 0.0161 & 1.26 $\times$ & \grey{109s} & 0.0150 & 1.35 $\times$ & \grey{84s} & \textbf{0.0146} & 1.38 $\times$ & \grey{90s} \\
        \myrowcolour
        \textsc{Breakfast Room} & 0.0094 & \grey{48s} & 0.0069 & 1.0 $\times$ & \grey{100s} & 0.0047 & 1.46 $\times$  & \grey{103s} & 0.0038 & 1.80 $\times$ & \grey{63s} & \textbf{0.0035} & 1.96 $\times$ & \grey{71s} \\
        
        \textsc{Living Room} & 0.0273 & \grey{32s} & 0.0184 & 1.0 $\times$ & \grey{74s} & 0.0146 & 1.26 $\times$ & \grey{80s} & 0.0157 & 1.17 $\times$ & \grey{47s} & \textbf{0.0132} & 1.39 $\times$ & \grey{54s} \\
        \myrowcolour
        \textsc{Pink Room} & 0.0046 & \grey{37s} & 0.0082 & 1.0 $\times$ & \grey{74s} & 0.0061 & 1.34 $\times$ & \grey{76s} & 0.0033 & 2.42 $\times$ & \grey{53s} & \textbf{0.0026} & 3.21 $\times$ & \grey{62s} \\
        
        \textsc{Salle de Bain} & 0.0819 & \grey{38s} & 0.0223 & 1.0 $\times$ & \grey{116s} & 0.0346 & 0.64 $\times$ & \grey{116s} & 0.0196 & 1.14 $\times$ & \grey{79s} & \textbf{0.0140} & 1.59 $\times$ & \grey{86s} \\ 
        \myrowcolour
        \textsc{Staircase} & 0.1812 & \grey{34s} & 0.0298 & 1.0 $\times$ & \grey{80s} & 0.0261 & 1.14 $\times$ & \grey{86s} & 0.0194 & 1.54 $\times$ & \grey{72s} & \textbf{0.0172} & 1.74 $\times$ & \grey{76s} \\
        
        \textsc{Veach Door} & 0.6208 & \grey{33s} & 0.2167 & 1.0 $\times$ & \grey{82s} & 0.1945 & 1.11 $\times$ & \grey{91s} & 0.0750 & 2.89 $\times$ & \grey{65s} & \textbf{0.0461} & 4.69 $\times$ & \grey{77s} \\      
        \myrowcolour
        \textsc{Veach Egg} & 8.2918 & \grey{33s} & 0.8379 & 1.0 $\times$ & \grey{82s} & 0.7870 & 1.07 $\times$ & \grey{85s} & 0.5984 & 1.40 $\times$ & \grey{62s} & \textbf{0.5352} & 1.56 $\times$ & \grey{69s} \\ 
        \textsc{White Room} & 0.0301 & \grey{38s} & 0.0278 & 1.0 $\times$ & \grey{107s} & 0.0253 & 1.10 $\times$ & \grey{103s} & 0.0124 & 2.25 $\times$ & \grey{76s} & \textbf{0.0100} & 2.75 $\times$ & \grey{87s} \\ 
        \bottomrule
    \end{tabular}
\end{table*}
}

\begin{figure*}[!ht]
  \centering
  \includegraphics[width=0.98\linewidth]{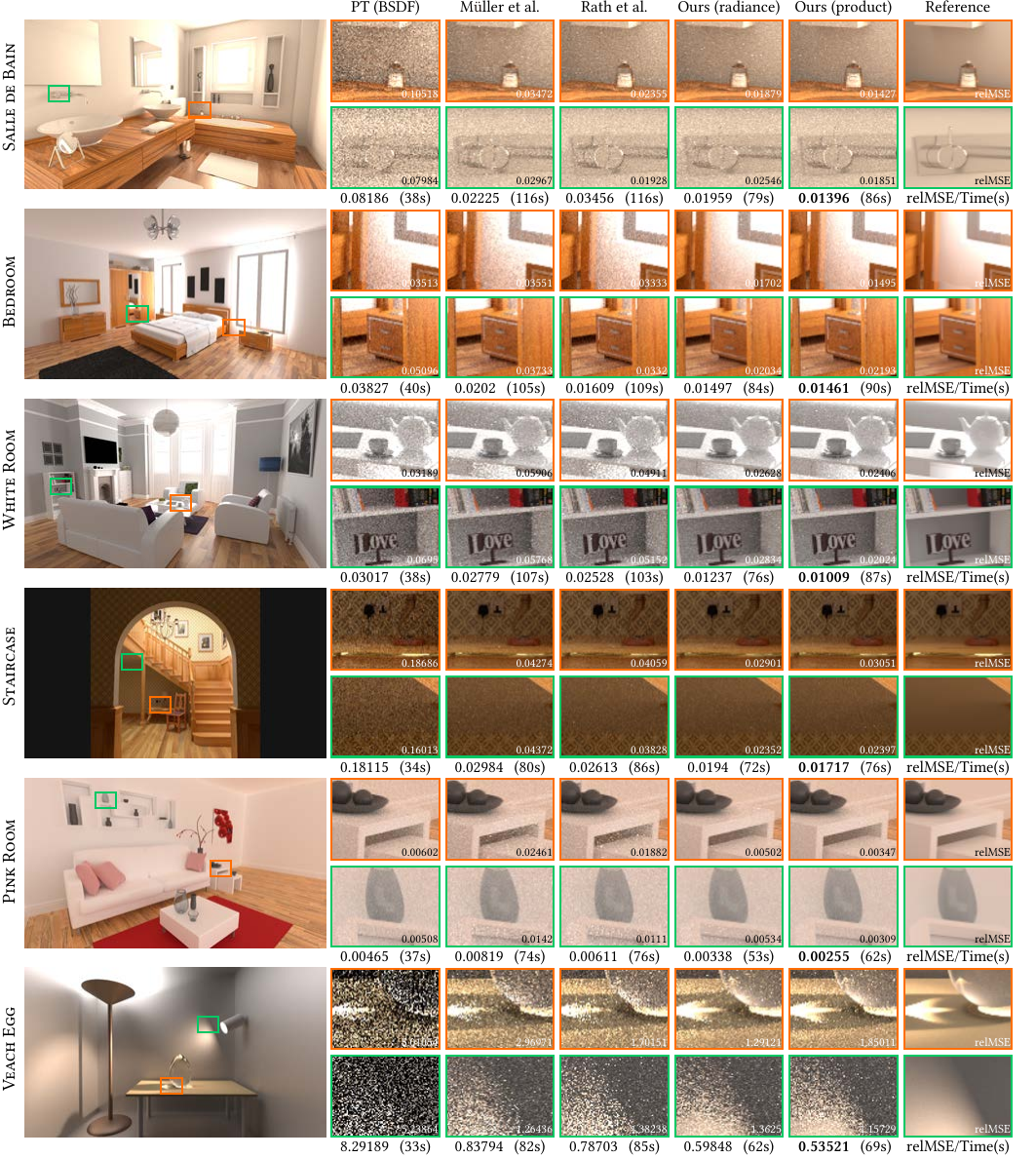}
  \caption{
  Visual comparisons using the same experimental setup with Fig. \ref{fig:core_comp}, all are rendered with 750spp at $1280\times 720$. We use the online training setup for all the guiding methods, i.e., all the samples are included in the final rendering. Our method exhibits better performance than other guiding methods in most scenes by only learning the incident radiance term while further reducing the error by incorporating the BSDF term (i.e., product sampling). More results on other test scenes, additional error metrics and false-color visualizations are provided in our supplementary interactive viewer.
  \label{fig:overall}} 
\end{figure*}


\begin{figure*}[!htb]
  \centering
  \includegraphics[width=0.98\linewidth]{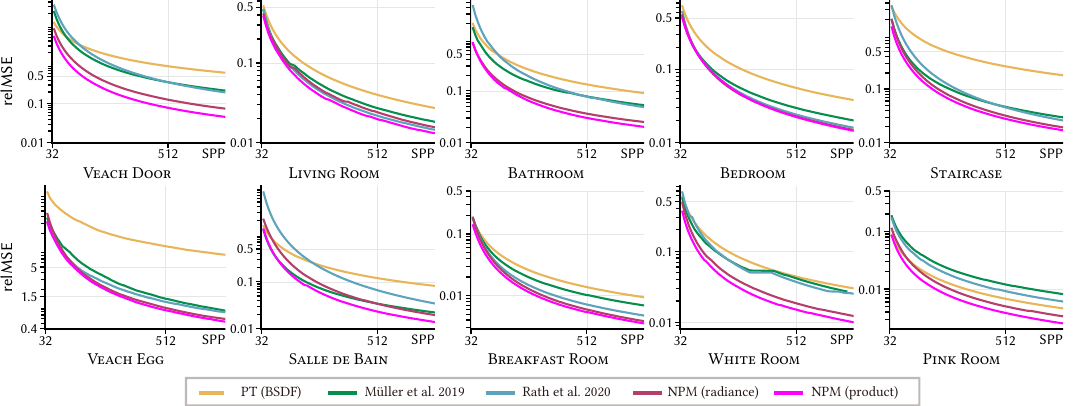}
  \caption{
    Convergence plots correspond to Fig.~\ref{fig:core_comp} and Fig.~\ref{fig:overall}.
    Unidirectional path tracing with BSDF importance sampling (PT-BSDF), Practical Path Guiding \cite{muller2019path}, Variance-aware Path Guiding \cite{rath2020variance} and our method with different target distributions ($\text{NPM-{radiance}}$ and $\text{NPM-{product}}$). Our methods consistently outperform these classical guiding methods, and quickly become effective even with a few training samples and short training time (e.g., 30spp, amounting to about 3 seconds on GPU), indicating practicality for preview or even interactive rendering. We attribute this success to the compact implicit representation and better spatial resolution of our method. The image results and detailed statistics could be inspected in the supplemental materials.
  \label{fig:convergence}}
\end{figure*}

\subsection{Comparisons}
\label{ssec:comparison}
Our method is compared against improved PPG \cite{muller2019path} (an enhanced version of Practical Path Guiding \cite{muller2017practical}), and Variance-aware Path Guiding~\cite{rath2020variance}. For the experimental configuration of the compared methods, we use the same as \cite{rath2020variance}, except for fixing the BSDF selection probability to $50\%$ (for both ours and the compared methods). Both compared methods used an iteratively reconstructed subdivision structure (i.e., the spatio-directional trees) to account for spatial variations. A total of 10 different scenes were tested.

We first show equal-spp comparisons on two representative scenes. The \textsc{Veach Door} scene features strong indirect illumination that is difficult to handle with BSDF importance sampling, while the \textsc{Bathroom} scene contains many specular and glossy surfaces.
As shown in Fig. \ref{fig:core_comp}, our proposed method outperforms the other two methods even when only learning incident radiance $L_{\mathrm{i}}$ ($\text{NPM-{radiance}}$). The noise is alleviated further with our full integrand learning method ($\text{NPM-{product}}$), since both of the scenes contain glossy surfaces, where the contribution of samples is strongly influenced by the BSDF term. We also note that our method quickly becomes effective at the very beginning of the training process (see the convergence plots in Fig.~\ref{fig:core_comp}). This indicates a better training efficiency over classical guiding methods, which will be discussed later. Additional results on more test scenes are shown in Fig.~\ref{fig:overall} and Tab. \ref{tab:comparisons}, as well as the convergence plots in Fig.~\ref{fig:convergence}.

\begin{figure}[b]
  \centering
  \includegraphics[width=\linewidth]{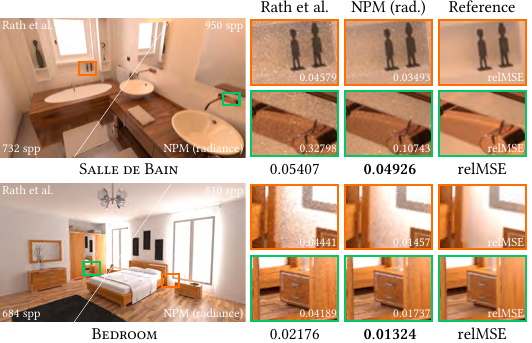}
  \caption{
    Equal-time comparisons (80s) on two test scenes between NPM(radiance) and Variance-aware Path Guiding \cite{rath2020variance}. 
  \label{fig:equaltime}}
\end{figure}

We then show the results of equal-time comparisons between our method and \cite{rath2020variance} in Fig.~\ref{fig:equaltime}.
Since they do not explicitly learn the product sampling distribution (i.e., conditioned on 5D inputs $\omega_{\mathrm{o}}$ and $\mathbf{x}$), we only use our radiance-based method (NPM-radiance) for fair comparisons.
Instead of simply learning the incident radiance distribution ($L_{\mathrm{i}}$), they use an improved target distribution to account for the variance and BSDF (marginalized over $\omega_{\mathrm{o}}$).
Our method, on the other hand, achieves better performance by learning $L_{\mathrm{i}}$ only. We attribute this superiority of our method to both the better capacity of capturing spatio-directional correlation and more parallelism. 

\begin{figure}[b]
  \centering
  \includegraphics[width=\linewidth]{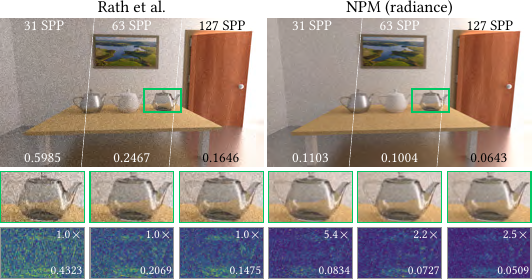}
  \caption{
  We train each guiding method with small training budgets (31 spp, 63 spp, 127 spp, respectively) and render the scene with 500 spp. Our method outperforms previous methods even with much fewer training samples.
  \label{fig:learning}}
\end{figure}

\begin{figure}[!htb]
  \centering
  \includegraphics[width=\linewidth]{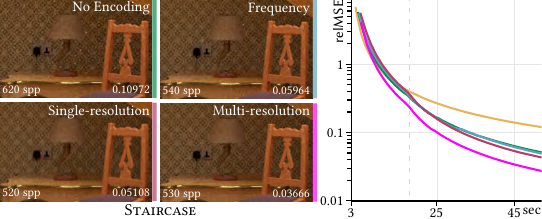}
  \caption{
  Equal-time comparison (50s) of different input encoding. We report the sample count and error (relMSE) of each method. The dashed line in the plot marks the end of the training phase. The multi-resolution spatial embedding outperforms other methods while remaining training-efficient. Yellow plot refers to path tracing with BSDF importance sampling.
  \label{fig:encoding}}
\end{figure}

\subsection{Evaluation}
\label{ssec:evaluation}

\paragraph{Trainable Spatial Embedding}
We analyze the performance of different forms of spatial input encoding in terms of convergence and quality (Fig. \ref{fig:encoding}). The spatial embedding (i.e. parametric encoding) uses trainable latent vector grids to model the spatially-varying target distributions, leaving the MLP to focus on decoding this implicit representation into valid vMF mixtures. The other two variants do not explicitly separate these two tasks by using a monolithic MLP.
The addition of spatial embedding significantly improves convergence, and the multi-resolution design further reduces error by better modeling finer-grained spatio-directional correlations. Furthermore, this does not introduce noticeable computational overhead, as only a small fraction of parameters are involved in each training/inference. 

\paragraph{Training Efficiency}
The effectiveness of guiding methods under small training budgets is important, especially for applications such as preview rendering or even interactive rendering.
We analyze the training efficiency of different guiding methods by comparing their performance under different training budgets (31 spp, 63 spp, 127 spp, respectively) in Fig.~\ref{fig:learning}.
Our method quickly converges to a good sampling distribution with only a few training samples and less training time cost (e.g., 31 spp with about 3s), thus outperforming previous guiding methods even with much fewer training samples.  

\subsection{Discussion}
\label{ssec:discussion}

\paragraph{Path Guiding Extensions}
Our method can be extended with many well-established extensions suggested by previous path guiding algorithms. They are straightforward to be integrated and are promising to further improve our performance. For example: (1) the BSDF selection probability could also be learned by our network or by some other caching strategies \cite{muller2020neural}, thus better handling the near-specular surfaces; and (2) the improved variance-aware target distribution \cite{rath2020variance} could be learned to account for the variance within the noisy MC estimates.

\paragraph{Performance Analysis}
Our method serves effective means for path guiding while remaining performance practical.
Specifically, the measured time cost per NPM evaluation (including both network inference and importance sampling the decoded mixture models) at $1280\times 720$ is about 3ms. Meanwhile, a training step (i.e., a batch of $2^{18}$ samples) costs about 10ms, indicating that a typical training process (about 1000 training steps) takes about 10s to converge on a single GPU. 
NPM contains a total of about $2\mathrm{M}$ learnable parameters, resulting in a memory consumption of $<10 \mathrm{MB}$. 
The compact design of our implicit NPM representation results in less control flow divergence, better memory locality, and better caching performance. Together, this makes our method practical for modern GPU parallelization, which is often harder to achieve with the tree-like spatial subdivision schemes used by most of the previous guiding methods.

\paragraph{Alternative Solutions}
Several studies also aim to tackle the parallax issue. Dodik et al. \shortcite{dodik2022path} use spatio-directional mixtures (i.e., conditioned on $\mathbf{x}$ and $\omega_{\mathrm{o}}$) to correlate target distributions with spatial positions. Ruppert et al. \shortcite{ruppert2020robust} design strategies to warp the guiding distributions in the spatial subdivisions to resemble the true distribution. However, these methods adopt sophisticated strategies that are difficult to parallelize efficiently on GPUs (e.g., batched expectation-maximization (EM) applied to a varying number of mixtures) while requiring extra efforts to fit scene BSDFs for product sampling.
In contrast, our method exploits trainable spatial embedding to encode the target distributions while using a decoder MLP to model the non-linearity between spatial features and PMMs in a GPU-friendly manner.
Nevertheless, incorporating ideas from these studies, such as adaptively controlling the granularity of learned distributions, may further enhance our method.

\section{Conclusion, Limitations and Future Work}
\label{sec:conclusion}

We present Neural Parametric Mixtures, a novel method for learning the target distributions for path guiding techniques. We use a compact implicit neural representation to encode the spatio-directional parametric distributions. Compared to previous non-neural methods that use explicit spatial subdivision structures to store directional distributions, our continuous implicit representation is simpler and more efficient while naturally avoiding the artifacts (e.g., parallax) caused by their discretized subdivision schemes. Our NPM technique could be efficiently trained with stochastic gradient descent to minimize the divergence from the target distribution. 

Despite the simplicity and effectiveness of our method, the main limitation resides in the lack of flexibility of our directional distribution representation, i.e., a fixed number of vMF components. While a similar issue exists in classical methods using PMMs \cite{dodik2022path, herholz2016product}, recent methods achieve more accurate directional distributions by adaptively merging and splitting the vMF components \cite{ruppert2020robust}. This, however, is non-trivial to apply to our NPM technique.

In future work, we will investigate more accurate approaches to implicitly encode parametric distributions while remaining efficient. Finding better basis functions or adaptively controlling the number of output components are two possible but challenging directions. Meanwhile, we would like to improve the efficiency of our method by using either novel architectural designs for neural networks, optimized implementation, or adapting previous extensions to path guiding algorithms. We believe these are important steps to make our method more practical for interactive or even real-time rendering pipelines, as well as other related applications that require fitting distributions with high-frequency spatial variations. In addition, applying our method to bidirectional path tracing \cite{PCBPT}, especially subspace probabilistic connections~\cite{Su2022TOG}, will also be an interesting future avenue.

\begin{acks}
This project was supported by the National Key R\&D Program of China (No.2022YFB3303400) and~\grantsponsor{nsfc}{NSFC of China}{} (No.~\grantnum{nsfc}{62172013}). We also thank the test scenes providers: Mareck (\textsc{Bathroom}), SlykDrako (\textsc{Bedroom}), Wig42 (\textsc{Breakfast Room}, \textsc{Living Room}, \textsc{Pink Room}, \textsc{Staircase}), nacimus (\textsc{Salle de Bain}), Jaakko Lehtinen (\textsc{Veach Door}), Jay-Artist (\textsc{White Room}), as well as the efforts for converting scene formats by Benedikt Bitterli \shortcite{resources16}.
\end{acks}

\bibliographystyle{ACM-Reference-Format}
\bibliography{references}


\begin{thebibliography}{34}


\ifx \showCODEN    \undefined \def \showCODEN     #1{\unskip}     \fi
\ifx \showDOI      \undefined \def \showDOI       #1{#1}\fi
\ifx \showISBNx    \undefined \def \showISBNx     #1{\unskip}     \fi
\ifx \showISBNxiii \undefined \def \showISBNxiii  #1{\unskip}     \fi
\ifx \showISSN     \undefined \def \showISSN      #1{\unskip}     \fi
\ifx \showLCCN     \undefined \def \showLCCN      #1{\unskip}     \fi
\ifx \shownote     \undefined \def \shownote      #1{#1}          \fi
\ifx \showarticletitle \undefined \def \showarticletitle #1{#1}   \fi
\ifx \showURL      \undefined \def \showURL       {\relax}        \fi
\providecommand\bibfield[2]{#2}
\providecommand\bibinfo[2]{#2}
\providecommand\natexlab[1]{#1}
\providecommand\showeprint[2][]{arXiv:#2}

\bibitem[Bitterli(2016)]%
        {resources16}
\bibfield{author}{\bibinfo{person}{Benedikt Bitterli}.}
  \bibinfo{year}{2016}\natexlab{}.
\newblock \bibinfo{title}{Rendering resources}.
\newblock
\newblock
\newblock
\shownote{https://benedikt-bitterli.me/resources/}.


\bibitem[Bus and Boubekeur(2017)]%
        {Bus2017DoubleHierarchies}
\bibfield{author}{\bibinfo{person}{Norbert Bus} {and} \bibinfo{person}{Tamy
  Boubekeur}.} \bibinfo{year}{2017}\natexlab{}.
\newblock \showarticletitle{Double Hierarchies for Directional Importance
  Sampling in Monte Carlo Rendering}.
\newblock \bibinfo{journal}{\emph{Journal of Computer Graphics Techniques
  (JCGT)}} \bibinfo{volume}{6}, \bibinfo{number}{3} (\bibinfo{date}{28 August}
  \bibinfo{year}{2017}), \bibinfo{pages}{25--37}.
\newblock
\showISSN{2331-7418}
\urldef\tempurl%
\url{http://jcgt.org/published/0006/03/02}
\showURL{%
\tempurl}


\bibitem[Currius et~al\mbox{.}(2020)]%
        {currius2020spherical}
\bibfield{author}{\bibinfo{person}{R.~R. Currius}, \bibinfo{person}{D.
  Dolonius}, \bibinfo{person}{U. Assarsson}, {and} \bibinfo{person}{E.
  Sintorn}.} \bibinfo{year}{2020}\natexlab{}.
\newblock \showarticletitle{Spherical Gaussian Light-field Textures for Fast
  Precomputed Global Illumination}.
\newblock \bibinfo{journal}{\emph{Computer Graphics Forum}}
  \bibinfo{volume}{39}, \bibinfo{number}{2} (\bibinfo{year}{2020}),
  \bibinfo{pages}{133--146}.
\newblock


\bibitem[Dinh et~al\mbox{.}(2017)]%
        {dinh2016density}
\bibfield{author}{\bibinfo{person}{Laurent Dinh}, \bibinfo{person}{Jascha
  Sohl-Dickstein}, {and} \bibinfo{person}{Samy Bengio}.}
  \bibinfo{year}{2017}\natexlab{}.
\newblock \showarticletitle{Density estimation using Real {NVP}}. In
  \bibinfo{booktitle}{\emph{International Conference on Learning
  Representations}}.
\newblock


\bibitem[Diolatzis et~al\mbox{.}(2022)]%
        {diolatzis2022active}
\bibfield{author}{\bibinfo{person}{Stavros Diolatzis}, \bibinfo{person}{Julien
  Philip}, {and} \bibinfo{person}{George Drettakis}.}
  \bibinfo{year}{2022}\natexlab{}.
\newblock \showarticletitle{Active Exploration for Neural Global Illumination
  of Variable Scenes}.
\newblock \bibinfo{journal}{\emph{ACM Transactions on Graphics}}
  (\bibinfo{year}{2022}).
\newblock


\bibitem[Dodik et~al\mbox{.}(2022)]%
        {dodik2022path}
\bibfield{author}{\bibinfo{person}{Ana Dodik}, \bibinfo{person}{Marios Papas},
  \bibinfo{person}{Cengiz {\"O}ztireli}, {and} \bibinfo{person}{Thomas
  M{\"u}ller}.} \bibinfo{year}{2022}\natexlab{}.
\newblock \showarticletitle{Path Guiding Using Spatio-Directional Mixture
  Models}. In \bibinfo{booktitle}{\emph{Computer Graphics Forum}},
  Vol.~\bibinfo{volume}{41}. Wiley Online Library, \bibinfo{pages}{172--189}.
\newblock


\bibitem[Hadadan et~al\mbox{.}(2021)]%
        {hadadan2021neural}
\bibfield{author}{\bibinfo{person}{Saeed Hadadan}, \bibinfo{person}{Shuhong
  Chen}, {and} \bibinfo{person}{Matthias Zwicker}.}
  \bibinfo{year}{2021}\natexlab{}.
\newblock \showarticletitle{Neural radiosity}.
\newblock \bibinfo{journal}{\emph{ACM Transactions on Graphics (TOG)}}
  \bibinfo{volume}{40}, \bibinfo{number}{6} (\bibinfo{year}{2021}),
  \bibinfo{pages}{1--11}.
\newblock


\bibitem[Herholz et~al\mbox{.}(2016)]%
        {herholz2016product}
\bibfield{author}{\bibinfo{person}{Sebastian Herholz}, \bibinfo{person}{Oskar
  Elek}, \bibinfo{person}{Ji{\v{r}}{\'\i} Vorba}, \bibinfo{person}{Hendrik
  Lensch}, {and} \bibinfo{person}{Jaroslav K{\v{r}}iv{\'a}nek}.}
  \bibinfo{year}{2016}\natexlab{}.
\newblock \showarticletitle{Product importance sampling for light transport
  path guiding}. In \bibinfo{booktitle}{\emph{Computer Graphics Forum}},
  Vol.~\bibinfo{volume}{35}. Wiley Online Library, \bibinfo{pages}{67--77}.
\newblock


\bibitem[Huo et~al\mbox{.}(2020)]%
        {huo2020adaptive}
\bibfield{author}{\bibinfo{person}{Yuchi Huo}, \bibinfo{person}{Rui Wang},
  \bibinfo{person}{Ruzahng Zheng}, \bibinfo{person}{Hualin Xu},
  \bibinfo{person}{Hujun Bao}, {and} \bibinfo{person}{Sung-Eui Yoon}.}
  \bibinfo{year}{2020}\natexlab{}.
\newblock \showarticletitle{Adaptive incident radiance field sampling and
  reconstruction using deep reinforcement learning}.
\newblock \bibinfo{journal}{\emph{ACM Transactions on Graphics (TOG)}}
  \bibinfo{volume}{39}, \bibinfo{number}{1} (\bibinfo{year}{2020}),
  \bibinfo{pages}{1--17}.
\newblock


\bibitem[Jakob(2012)]%
        {jakob2012numerically}
\bibfield{author}{\bibinfo{person}{Wenzel Jakob}.}
  \bibinfo{year}{2012}\natexlab{}.
\newblock \showarticletitle{Numerically stable sampling of the von Mises-Fisher
  distribution on Sˆ2 (and other tricks)}.
\newblock \bibinfo{journal}{\emph{Interactive Geometry Lab, ETH Z{\"u}rich,
  Tech. Rep}} (\bibinfo{year}{2012}), \bibinfo{pages}{6}.
\newblock


\bibitem[Jensen(1995)]%
        {jensen1995importance}
\bibfield{author}{\bibinfo{person}{Henrik~Wann Jensen}.}
  \bibinfo{year}{1995}\natexlab{}.
\newblock \showarticletitle{Importance driven path tracing using the photon
  map}. In \bibinfo{booktitle}{\emph{Eurographics Workshop on Rendering
  Techniques}}. Springer, \bibinfo{pages}{326--335}.
\newblock


\bibitem[Kajiya(1986)]%
        {kaj86re}
\bibfield{author}{\bibinfo{person}{James~T. Kajiya}.}
  \bibinfo{year}{1986}\natexlab{}.
\newblock \showarticletitle{The Rendering Equation}.
\newblock \bibinfo{journal}{\emph{SIGGRAPH Comput. Graph.}}
  (\bibinfo{year}{1986}).
\newblock


\bibitem[Kingma and Ba(2015)]%
        {kingma2014adam}
\bibfield{author}{\bibinfo{person}{Diederik~P. Kingma} {and}
  \bibinfo{person}{Jimmy Ba}.} \bibinfo{year}{2015}\natexlab{}.
\newblock \showarticletitle{Adam: A Method for Stochastic Optimization}.
\newblock \bibinfo{journal}{\emph{ICLR}} (\bibinfo{year}{2015}).
\newblock


\bibitem[Lafortune and Willems(1995)]%
        {lafortune19955d}
\bibfield{author}{\bibinfo{person}{Eric~P Lafortune} {and}
  \bibinfo{person}{Yves~D Willems}.} \bibinfo{year}{1995}\natexlab{}.
\newblock \showarticletitle{A 5D tree to reduce the variance of Monte Carlo ray
  tracing}. In \bibinfo{booktitle}{\emph{Eurographics Workshop on Rendering
  Techniques}}. Springer, \bibinfo{pages}{11--20}.
\newblock


\bibitem[Laine et~al\mbox{.}(2013)]%
        {laine2013megakernels}
\bibfield{author}{\bibinfo{person}{Samuli Laine}, \bibinfo{person}{Tero
  Karras}, {and} \bibinfo{person}{Timo Aila}.} \bibinfo{year}{2013}\natexlab{}.
\newblock \showarticletitle{Megakernels considered harmful: Wavefront path
  tracing on GPUs}. In \bibinfo{booktitle}{\emph{Proceedings of the 5th
  High-Performance Graphics Conference}}. \bibinfo{pages}{137--143}.
\newblock


\bibitem[Mildenhall et~al\mbox{.}(2020)]%
        {mildenhall2020nerf}
\bibfield{author}{\bibinfo{person}{Ben Mildenhall}, \bibinfo{person}{Pratul~P.
  Srinivasan}, \bibinfo{person}{Matthew Tancik}, \bibinfo{person}{Jonathan~T.
  Barron}, \bibinfo{person}{Ravi Ramamoorthi}, {and} \bibinfo{person}{Ren Ng}.}
  \bibinfo{year}{2020}\natexlab{}.
\newblock \showarticletitle{NeRF: Representing Scenes as Neural Radiance Fields
  for View Synthesis}. In \bibinfo{booktitle}{\emph{ECCV}}.
\newblock


\bibitem[M\"{u}ller(2019)]%
        {muller2019path}
\bibfield{author}{\bibinfo{person}{Thomas M\"{u}ller}.}
  \bibinfo{year}{2019}\natexlab{}.
\newblock \showarticletitle{"Practical Path Guiding" in Production}. In
  \bibinfo{booktitle}{\emph{ACM SIGGRAPH 2019 Courses}}
  \emph{(\bibinfo{series}{SIGGRAPH '19})}. \bibinfo{publisher}{ACM},
  \bibinfo{address}{New York, NY, USA}, Article \bibinfo{articleno}{18},
  \bibinfo{numpages}{77}~pages.
\newblock
\showISBNx{978-1-4503-6307-5}


\bibitem[M\"uller et~al\mbox{.}(2022)]%
        {muller2022instant}
\bibfield{author}{\bibinfo{person}{Thomas M\"uller}, \bibinfo{person}{Alex
  Evans}, \bibinfo{person}{Christoph Schied}, {and} \bibinfo{person}{Alexander
  Keller}.} \bibinfo{year}{2022}\natexlab{}.
\newblock \showarticletitle{Instant Neural Graphics Primitives with a
  Multiresolution Hash Encoding}.
\newblock \bibinfo{journal}{\emph{ACM Trans. Graph.}} \bibinfo{volume}{41},
  \bibinfo{number}{4}, Article \bibinfo{articleno}{102} (\bibinfo{date}{July}
  \bibinfo{year}{2022}), \bibinfo{numpages}{15}~pages.
\newblock


\bibitem[M{\"u}ller et~al\mbox{.}(2017)]%
        {muller2017practical}
\bibfield{author}{\bibinfo{person}{Thomas M{\"u}ller}, \bibinfo{person}{Markus
  Gross}, {and} \bibinfo{person}{Jan Nov{\'a}k}.}
  \bibinfo{year}{2017}\natexlab{}.
\newblock \showarticletitle{Practical path guiding for efficient
  light-transport simulation}. In \bibinfo{booktitle}{\emph{Computer Graphics
  Forum}}, Vol.~\bibinfo{volume}{36}. Wiley Online Library,
  \bibinfo{pages}{91--100}.
\newblock


\bibitem[M{\"u}ller et~al\mbox{.}(2019)]%
        {muller2019neural}
\bibfield{author}{\bibinfo{person}{Thomas M{\"u}ller}, \bibinfo{person}{Brian
  McWilliams}, \bibinfo{person}{Fabrice Rousselle}, \bibinfo{person}{Markus
  Gross}, {and} \bibinfo{person}{Jan Nov{\'a}k}.}
  \bibinfo{year}{2019}\natexlab{}.
\newblock \showarticletitle{Neural importance sampling}.
\newblock \bibinfo{journal}{\emph{ACM Transactions on Graphics (TOG)}}
  \bibinfo{volume}{38}, \bibinfo{number}{5} (\bibinfo{year}{2019}),
  \bibinfo{pages}{1--19}.
\newblock


\bibitem[M{\"u}ller et~al\mbox{.}(2020)]%
        {muller2020neural}
\bibfield{author}{\bibinfo{person}{Thomas M{\"u}ller}, \bibinfo{person}{Fabrice
  Rousselle}, \bibinfo{person}{Alexander Keller}, {and} \bibinfo{person}{Jan
  Nov{\'a}k}.} \bibinfo{year}{2020}\natexlab{}.
\newblock \showarticletitle{Neural control variates}.
\newblock \bibinfo{journal}{\emph{ACM Transactions on Graphics (TOG)}}
  \bibinfo{volume}{39}, \bibinfo{number}{6} (\bibinfo{year}{2020}),
  \bibinfo{pages}{1--19}.
\newblock


\bibitem[M\"{u}ller et~al\mbox{.}(2021)]%
        {muller2021real}
\bibfield{author}{\bibinfo{person}{Thomas M\"{u}ller}, \bibinfo{person}{Fabrice
  Rousselle}, \bibinfo{person}{Jan Nov\'{a}k}, {and} \bibinfo{person}{Alexander
  Keller}.} \bibinfo{year}{2021}\natexlab{}.
\newblock \showarticletitle{Real-Time Neural Radiance Caching for Path
  Tracing}.
\newblock \bibinfo{journal}{\emph{ACM Trans. Graph.}} \bibinfo{volume}{40},
  \bibinfo{number}{4}, Article \bibinfo{articleno}{36} (\bibinfo{date}{jul}
  \bibinfo{year}{2021}), \bibinfo{numpages}{16}~pages.
\newblock
\showISSN{0730-0301}


\bibitem[Munkberg et~al\mbox{.}(2022)]%
        {munkberg2021nvdiffrec}
\bibfield{author}{\bibinfo{person}{Jacob Munkberg}, \bibinfo{person}{Jon
  Hasselgren}, \bibinfo{person}{Tianchang Shen}, \bibinfo{person}{Jun Gao},
  \bibinfo{person}{Wenzheng Chen}, \bibinfo{person}{Alex Evans},
  \bibinfo{person}{Thomas Mueller}, {and} \bibinfo{person}{Sanja Fidler}.}
  \bibinfo{year}{2022}\natexlab{}.
\newblock \showarticletitle{{Extracting Triangular 3D Models, Materials, and
  Lighting From Images}}.
\newblock \bibinfo{journal}{\emph{CVPR}} (\bibinfo{year}{2022}).
\newblock


\bibitem[Müller(2021)]%
        {muller2021tcnn}
\bibfield{author}{\bibinfo{person}{Thomas Müller}.}
  \bibinfo{year}{2021}\natexlab{}.
\newblock \bibinfo{booktitle}{\emph{{tiny-cuda-nn}}}.
\newblock
\urldef\tempurl%
\url{https://github.com/NVlabs/tiny-cuda-nn}
\showURL{%
\tempurl}


\bibitem[Parker et~al\mbox{.}(2010)]%
        {parker2010optix}
\bibfield{author}{\bibinfo{person}{Steven~G Parker}, \bibinfo{person}{James
  Bigler}, \bibinfo{person}{Andreas Dietrich}, \bibinfo{person}{Heiko
  Friedrich}, \bibinfo{person}{Jared Hoberock}, \bibinfo{person}{David Luebke},
  \bibinfo{person}{David McAllister}, \bibinfo{person}{Morgan McGuire},
  \bibinfo{person}{Keith Morley}, \bibinfo{person}{Austin Robison},
  {et~al\mbox{.}}} \bibinfo{year}{2010}\natexlab{}.
\newblock \showarticletitle{Optix: a general purpose ray tracing engine}.
\newblock \bibinfo{journal}{\emph{ACM Transactions on Graphics (TOG)}}
  \bibinfo{volume}{29}, \bibinfo{number}{4} (\bibinfo{year}{2010}),
  \bibinfo{pages}{1--13}.
\newblock


\bibitem[Popov et~al\mbox{.}(2015)]%
        {PCBPT}
\bibfield{author}{\bibinfo{person}{S. Popov}, \bibinfo{person}{R. Ramamoorthi},
  \bibinfo{person}{F. Durand}, {and} \bibinfo{person}{G. Drettakis}.}
  \bibinfo{year}{2015}\natexlab{}.
\newblock \showarticletitle{Probabilistic Connections for Bidirectional Path
  Tracing}.
\newblock \bibinfo{journal}{\emph{Computer Graphics Forum}}
  \bibinfo{volume}{34}, \bibinfo{number}{4} (\bibinfo{date}{07}
  \bibinfo{year}{2015}), \bibinfo{pages}{75--86}.
\newblock


\bibitem[Rath et~al\mbox{.}(2020)]%
        {rath2020variance}
\bibfield{author}{\bibinfo{person}{Alexander Rath}, \bibinfo{person}{Pascal
  Grittmann}, \bibinfo{person}{Sebastian Herholz}, \bibinfo{person}{Petr
  V{\'e}voda}, \bibinfo{person}{Philipp Slusallek}, {and}
  \bibinfo{person}{Jaroslav K{\v{r}}iv{\'a}nek}.}
  \bibinfo{year}{2020}\natexlab{}.
\newblock \showarticletitle{Variance-aware path guiding}.
\newblock \bibinfo{journal}{\emph{ACM Transactions on Graphics (TOG)}}
  \bibinfo{volume}{39}, \bibinfo{number}{4} (\bibinfo{year}{2020}),
  \bibinfo{pages}{151--1}.
\newblock


\bibitem[Ruppert et~al\mbox{.}(2020)]%
        {ruppert2020robust}
\bibfield{author}{\bibinfo{person}{Lukas Ruppert}, \bibinfo{person}{Sebastian
  Herholz}, {and} \bibinfo{person}{Hendrik~PA Lensch}.}
  \bibinfo{year}{2020}\natexlab{}.
\newblock \showarticletitle{Robust fitting of parallax-aware mixtures for path
  guiding}.
\newblock \bibinfo{journal}{\emph{ACM Transactions on Graphics (TOG)}}
  \bibinfo{volume}{39}, \bibinfo{number}{4} (\bibinfo{year}{2020}),
  \bibinfo{pages}{147--1}.
\newblock


\bibitem[Su et~al\mbox{.}(2022)]%
        {Su2022TOG}
\bibfield{author}{\bibinfo{person}{Fujia Su}, \bibinfo{person}{Sheng Li}, {and}
  \bibinfo{person}{Guoping Wang}.} \bibinfo{year}{2022}\natexlab{}.
\newblock \showarticletitle{SPCBPT: Subspace-Based Probabilistic Connections
  for Bidirectional Path Tracing}.
\newblock \bibinfo{journal}{\emph{ACM Trans. Graph.}} \bibinfo{volume}{41},
  \bibinfo{number}{4}, Article \bibinfo{articleno}{77} (\bibinfo{date}{jul}
  \bibinfo{year}{2022}), \bibinfo{numpages}{14}~pages.
\newblock
\showISSN{0730-0301}
\urldef\tempurl%
\url{https://doi.org/10.1145/3528223.3530183}
\showDOI{\tempurl}


\bibitem[Verbin et~al\mbox{.}(2022)]%
        {verbin2022refnerf}
\bibfield{author}{\bibinfo{person}{Dor Verbin}, \bibinfo{person}{Peter Hedman},
  \bibinfo{person}{Ben Mildenhall}, \bibinfo{person}{Todd Zickler},
  \bibinfo{person}{Jonathan~T. Barron}, {and} \bibinfo{person}{Pratul~P.
  Srinivasan}.} \bibinfo{year}{2022}\natexlab{}.
\newblock \showarticletitle{{Ref-NeRF}: Structured View-Dependent Appearance
  for Neural Radiance Fields}.
\newblock \bibinfo{journal}{\emph{CVPR}} (\bibinfo{year}{2022}).
\newblock


\bibitem[Vorba et~al\mbox{.}(2019)]%
        {vorba2019production}
\bibfield{author}{\bibinfo{person}{Ji\v{r}\'{\i} Vorba},
  \bibinfo{person}{Johannes Hanika}, \bibinfo{person}{Sebastian Herholz},
  \bibinfo{person}{Thomas M\"{u}ller}, \bibinfo{person}{Jaroslav
  K\v{r}iv\'{a}nek}, {and} \bibinfo{person}{Alexander Keller}.}
  \bibinfo{year}{2019}\natexlab{}.
\newblock \showarticletitle{Path Guiding in Production}. In
  \bibinfo{booktitle}{\emph{ACM SIGGRAPH 2019 Courses}} (Los Angeles,
  California) \emph{(\bibinfo{series}{SIGGRAPH '19})}.
  \bibinfo{publisher}{ACM}, \bibinfo{address}{New York, NY, USA}, Article
  \bibinfo{articleno}{18}, \bibinfo{numpages}{77}~pages.
\newblock
\showISBNx{978-1-4503-6307-5}


\bibitem[Vorba et~al\mbox{.}(2014)]%
        {vorba2014line}
\bibfield{author}{\bibinfo{person}{Ji{\v{r}}{\'\i} Vorba},
  \bibinfo{person}{Ond{\v{r}}ej Karl{\'\i}k}, \bibinfo{person}{Martin
  {\v{S}}ik}, \bibinfo{person}{Tobias Ritschel}, {and}
  \bibinfo{person}{Jaroslav K{\v{r}}iv{\'a}nek}.}
  \bibinfo{year}{2014}\natexlab{}.
\newblock \showarticletitle{On-line learning of parametric mixture models for
  light transport simulation}.
\newblock \bibinfo{journal}{\emph{ACM Transactions on Graphics (TOG)}}
  \bibinfo{volume}{33}, \bibinfo{number}{4} (\bibinfo{year}{2014}),
  \bibinfo{pages}{1--11}.
\newblock


\bibitem[Yu et~al\mbox{.}(2021)]%
        {yu2021plenoctrees}
\bibfield{author}{\bibinfo{person}{Alex Yu}, \bibinfo{person}{Ruilong Li},
  \bibinfo{person}{Matthew Tancik}, \bibinfo{person}{Hao Li},
  \bibinfo{person}{Ren Ng}, {and} \bibinfo{person}{Angjoo Kanazawa}.}
  \bibinfo{year}{2021}\natexlab{}.
\newblock \showarticletitle{{PlenOctrees} for Real-time Rendering of Neural
  Radiance Fields}. In \bibinfo{booktitle}{\emph{ICCV}}.
\newblock


\bibitem[Zhu et~al\mbox{.}(2021)]%
        {zhu2021hierarchical}
\bibfield{author}{\bibinfo{person}{Shilin Zhu}, \bibinfo{person}{Zexiang Xu},
  \bibinfo{person}{Tiancheng Sun}, \bibinfo{person}{Alexandr Kuznetsov},
  \bibinfo{person}{Mark Meyer}, \bibinfo{person}{Henrik~Wann Jensen},
  \bibinfo{person}{Hao Su}, {and} \bibinfo{person}{Ravi Ramamoorthi}.}
  \bibinfo{year}{2021}\natexlab{}.
\newblock \showarticletitle{Hierarchical neural reconstruction for path guiding
  using hybrid path and photon samples}.
\newblock \bibinfo{journal}{\emph{ACM Transactions on Graphics (TOG)}}
  \bibinfo{volume}{40}, \bibinfo{number}{4} (\bibinfo{year}{2021}),
  \bibinfo{pages}{1--16}.
\newblock


\end{thebibliography}


\end{document}